\documentclass[journal=jctc,manuscript=article]{achemso}
\setkeys{acs}{articletitle = true}
\usepackage{hyperref}

\usepackage{blindtext}

\usepackage{changes}
\usepackage{lipsum}
\usepackage{graphicx}
\usepackage{caption}
\usepackage{subcaption}

\usepackage{bm}
\usepackage{amsmath}
\usepackage{hyperref} \hypersetup{colorlinks=true,citecolor=blue,linkcolor=blue,urlcolor=blue}
\usepackage{amsmath}
\usepackage{multirow}
\usepackage{booktabs}
\usepackage{epsfig}
\usepackage{xcolor,soul}
\usepackage{epstopdf}
\usepackage{lipsum,adjustbox}
\usepackage{tikz}      
\usetikzlibrary{matrix,
	chains,
	positioning,
	decorations.pathreplacing,
	arrows,calc,shapes, backgrounds,
	shapes,
	shapes.geometric,
	shapes.symbols,
	shapes.arrows,
	shapes.multipart,
	shapes.callouts,
	shapes.misc}
\tikzset{fontscale/.style = {font=\relsize{#1}}
}
\tikzstyle{block}=[draw opacity=0.7,line width=1.4cm]
\tikzstyle{stuff_fill}=[chamfered rectangle,draw,fill=white!30,font={C}]
\tikzstyle{circ_fill}=[circle,draw,fill=white!30,font={C}]
\tikzstyle{circ_fill_red}=[circle,draw,fill=red!30,font={C}]
\tikzstyle{block} = [draw, rectangle, text width=3em, text centered, minimum height=8mm,     node distance=2.3em]
\tikzstyle{line} = [draw]
\tikzstyle{Mult_fil}=[rounded rectangle, inner sep=1ex,draw,fill=blue!10,font={C}, minimum height=6ex, minimum width=12ex, font=\fontsize{14}{14}\selectfont]
\tikzset{MyArrow/.style={single arrow,font=\fontsize{12}{12}\selectfont, draw,fill=red!30, minimum width=4ex, minimum height=18ex, inner sep=1ex, single arrow head extend=2ex},
	block/.style={
		draw,
		rectangle, 
		text width=3em, 
		text centered, 
		minimum height=8mm,     
		node distance=2.3em
	}, 
	line/.style={draw}
}

\setstcolor{red}    
\bibpunct{[}{]}{,}{n}{}{}   
\newcommand{\dblquote}[1]{\textquotedblleft #1\textquotedblright}

\author{Natalia Kuritz}
\affiliation{Department of Physical Electronics, Tel-Aviv University, Tel-Aviv 69978, Israel}
\author{Goren Gordon}
\affiliation{Department of Industrial Engineering, Tel-Aviv University, Tel-Aviv 69978, Israel}
\author{Amir Natan}
\affiliation{Department of Physical Electronics, Tel-Aviv University, Tel-Aviv 69978, Israel}
\altaffiliation{The Sackler Center for Computational Molecular and Materials Science, Tel-Aviv University, Tel-Aviv 69978, Israel }
\email{amirnatan@post.tau.ac.il}

\title[An \textsf{achemso} demo]
{Size and Temperature Transferability of Direct and Local Deep Neural Networks for\\ Atomic Forces}

\abbreviations{MD, ML, DL, DNN, DFT}

\begin{document}

\begin{abstract}
A direct and local deep learning (DL) model for atomic forces is presented. We demonstrate the model performance in bulk aluminum, sodium, and silicon; and show that its errors are comparable to those found in state-of-the-art machine learning and DL models. We then analyze the model's performance as a function of the number of neighbors included and show that one can ascertain physical attributes of the system from the analysis of the deep learning model's behavior. Finally, we test the size scaling performance of the model, and the transferability between different temperatures, and show that our model performs well in both scaling to larger systems and high-to-low temperature predictability.
\end{abstract}

\maketitle

\section{Introduction}
The computation of large systems' atomistic dynamics is required in fields such as biochemistry, surface science, electrochemistry and many others. 
Ab-initio molecular dynamics (AIMD)~\cite{marx2012ab} is a powerful tool, but can have a high computational cost which prohibits the computation of large systems for long enough time intervals. 
A successful approach, with a significantly lower computational cost, is the use of classical force fields (FF) to model the forces between the atoms~\cite{allen1989computer,frenkel2001understanding}. This scheme enables the simulation of the dynamics of large systems (more than 10$^6$ atoms) within the nanosecond and microsecond timescales. The disadvantage of classical models is that they often need system-specific parametrization and cannot handle chemical reactions where molecules break or form new bonds. 
Another example that can be challenging for a classical FF approach is that of metal oxidation; a metal atom is neutral inside the metal bulk, but is charged inside the oxide layer. A possible way to model such varying environments is to introduce more complicated FF, such as the variable charge force field~\cite{Sen2015}, COMB~\cite{Liang} and ReaxFF~\cite{Liang2013,Bartok2013}. Such FF can successfully describe more challenging situations, but need greater parametrization and, again, cannot cover all possible atomic configurations.

An approach that was developed in the last decade is to use Machine Learning (ML) and Deep Learning (DL)~\cite{Goodfellow2015} algorithms to build \dblquote{on the fly} computationally cheap predictors for the energy, forces, and other physical properties. This approach enables the performance of calculations with an accuracy that is close enough to fully quantum molecular dynamics (MD), but with running speeds that are more than 100 times faster.

One way to tackle the statistical learning of chemical properties is kernel-based ML. Within this approach the atomic system is represented by physical fingerprints such as the \dblquote{Coulomb} matrix \cite{Rupp2012,Montavon2013b,Hansen2013b}, the bag of bonds (BoB)~\cite{Hansen2015}, scattering transforms~\cite{Hirn2015}, bispectrum~\cite{Bartok2010}, smooth overlap of atomic positions (SOAP)~\cite{Bartok2013}, generalized symmetry functions \cite{Behler2011}, bonding angular machine learning \cite{Huang2016}, tensor representation \cite{Huo2017}, Gaussian processes regression \cite{Li2015,Glielmo2017} and more \cite{VonLilienfeld2015,Schutt2014a,Schutt2017,Botu2015c,faber2016machine,mueller2016machine}.

Several DL implementations have been developed recently~\cite{Gastegger2017a,Goh,Behler2017b,Zhang}. It is possible to define two main DL algorithm strategies: \textit{convolutional neural network} (CNN) \cite{Goh,Goh2017a,Schutt2017} 
and \textit{fully connected deep neural network }(DNN) \cite{Behler2015b, Behler2014a, Gastegger2015, Gastegger2016}. The DNN approach is size-extensive and is the simplest option to model energies, as shown in several recent studies.

A desired goal for both ML and DL approaches is to be able to train the model on a small system, and then to use the accrued knowledge in much larger and diverse environments. This requires a formulation of a \dblquote{local environment} input to the model.

DL models with a local environment input were recently suggested by Han et al. \cite{Han} and by Lubbers et al.~\cite{Lubbers2017}. In these models, the input of the system is presented as a simple function of the atomic positions of each atom's neighbors. The output of the model was the energy, while the forces were estimated from the energy derivatives with respect to the atoms' locations.

In this work, we describe the construction and use of local environments for a DNN based model for the forces in solids. This model is very close in spirit to the one reported by Han et al.~\cite{Han}, but makes a direct prediction of the forces instead of the energy. We first show that with this DNN model we can reach an accuracy that is comparable to that found in state-of-the-art ML and DL models. We demonstrate this for bulk aluminum (Al), silicon (Si), and sodium (Na) at temperatures of 300K and 2000K. We then analyze the dependence of the error on the number of neighbors that are used for the input and show that physical attributes of the underlying system can be learned from this analysis. Finally, we analyze the ability of the developed DL model to do actual size scaling, that is, to use training in a small cell for prediction in larger cells. We also test the ability to operate at different temperatures, i.e. to train at one temperature and predict at another temperature. We conclude with a discussion of how to proceed and build fully scalable and transferable DL models that can work with a wide range of environments. The rest of the paper is organized as follows: we first describe the model, methods and data sets; we then show the results for same size, different size, and different temperature systems. Finally, we discuss the meaning of some of the observations and the challenges faced in developing a fully scalable prediction.

\section{Methods}
\subsection{Tools}
We used the VASP~\cite{Kresse1993, Kresse1994, Kresse1996a} package for all quantum simulations of bulk Al and bulk Na. We analyzed VASP files to get the radial distribution function (RDF) using the AFLOW~\cite{Curtarolo} package.
 We used the folowing python packages Python libraries for building the network structure and for the training: numpy and scipy~\cite{VanderWalt2011}, Pandas~\cite{McKinney2010}, IPython~\cite{Perez2007}, Matplotlib~\cite{Hunter2007}, tensorflow~\cite{Abadi2016}, tflearn~\cite{tflearn2016}, ASE~\cite{Larsen2017} and Scikit-Learn~\cite{Pedregosa2012}. All the code was written in Python. In addition, the Matlab software~\cite{MATLAB:2017} was used to draw most of the figures.

\subsection{Data sets}
We used the following bulk super-cells for the training and validation of Al and Na: a 3x3x3 super-cell (27 atoms), and 5x5x5 super-cell (125 atoms). For Si, we used super-cells of 2x2x2 (16 atoms) and 4x4x4 (128 atoms).

The training set consisted of about 1620 MD steps that we randomly chose from a trajectory of 1800 steps. The other 180 configurations were used for validation. When changing the super-cell size $M\times M\times M$, we changed the number of k-points, defined by a Monkhorst-Pack~\cite{Monkhorst1976} grid $M_k\times M_k\times M_k$, so that the Born-von Karman cell~\cite{ashcroft1976solid} stays roughly constant (i.e., $M\times M_k$ is kept constant) and so the level of electronic sampling is similar.

The data sets were prepared with the following protocol. First, the super-cells were built for each material:  Face-centered cubic (fcc) for Al, Body-centered cubic (bcc) for Na, and Diamond structure for Si with the experimental lattice constants 4.05\AA, 4.29\AA \, and 5.43\AA, \ respectively~\cite{wyckoff1963crystal}. Then an AIMD was run with VASP for Na and Si and classical MD for Al with the EMT force field and the ASE package. The MD was performed with a constant super-cell volume and shape to produce the relevant atomic positions. For Al, the atomic configuration was recorded each 50fs. For Na and Si, the atomic configuration was recorded each 1fs. The MD time propagation was performed in the canonical ensemble and Nos\`{e} algorithm for Si and Na. The Al run was conducted in the microcanonical ensemble with temperatures around T=300K and T=2000K. For the final calculation of the forces, we applied the following protocol. Each of the produced structures was run with Density Functional Theory (DFT)~\cite{koch2000chemist} without further geometrical relaxation, we used the PBE functional~\cite{Perdew1996} without spin-polarization, the VASP PAW pseudopotentials~\cite{Blochl1994,Kresse1999}, and an energy cutoff of 260eV for Si and Na, and 520eV for Al, which was found to be sufficient for the accuracies we report later. The number of k-points for each of the cells is shown in table \ref{tab:MAEComp}.

\subsection{Deep learning models and learning procedure}
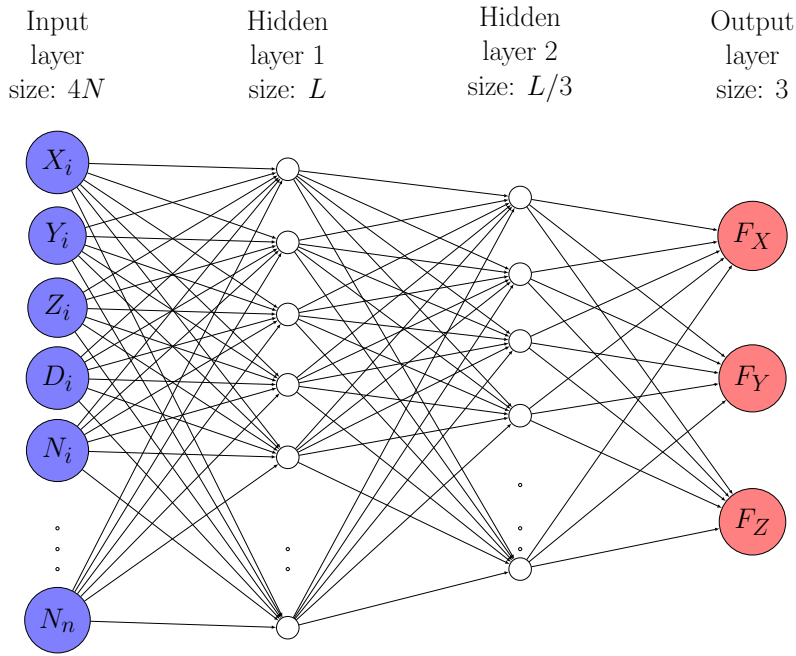
\begin{figure}[t]
	\centering
	\begin{adjustbox}{width=0.7\textwidth}
	\begin{tikzpicture}[
	plain/.style={
		draw=none,
		fill=none,
	},
	plainOut/.style={
		fill=red!50,
		font=\huge,
	},
plainETC/.style={
	inner sep=1pt,
	column sep=2cm,
	row sep=7pt
},
plainInp/.style={
	fill=blue!50,
	font=\Huge,
},
	net/.style={
		matrix of nodes,
		nodes={
			draw,
			circle,
			inner sep=7pt
		},
		nodes in empty cells,
		column sep=2cm,
		row sep=-7pt
	}, 
	>=latex
	]
	\matrix[net] (mat)
	{ 
		|[plain]| \parbox{3.5cm}{\centering \Huge Input\\layer\\size: $4N$} & |[plain]| \parbox{3.5cm}{\centering \Huge  Hidden\\layer 1\\size: $L$} & |[plain]|\parbox{3.5cm}{\centering \Huge    Hidden\\layer 2\\size: $L/3$}&|[plain]| \parbox{3.5cm}{\centering \Huge Output\\layer\\size: $3$} \\
		|[plainInp]|$X_i$ &         &|[plain]| &|[plain]| \\
		|[plain]|   &|[plain]|& \\
		|[plainInp]|$Y_i$ &         &|[plain]| &|[plainOut]|$F_X$\\
		|[plain]|   &|[plain]|& \\
		|[plainInp]|$Z_i$&         &|[plain]| &|[plain]|\\
		|[plain]|    &|[plain]|&\\
		|[plainInp]|$D_i$ &         &|[plain]| &|[plainOut]|$F_Y$\\
		|[plain]|    &|[plain]|&\\
		|[plainInp]|$N_i$ &         &|[plain]| &|[plain]| \\
		|[plain]|    &|[plain]|&|[plainETC]|\\
		|[plainETC]| &|[plain]|& |[plainETC]|&|[plainOut]|$F_Z$\\
		|[plainETC]| &|[plainETC]|&|[plainETC]| \\
		|[plain]|    &|[plain]|\\
		|[plainETC]| &|[plainETC]|& \\
		|[plain]|    &|[plain]|\\
		|[plainInp]|$N_n$&              \\
	};
	\foreach \ai in {2,4,...,10,17}
	{\foreach \aii in {2,4,...,10,17}
		\draw[->] (mat-\ai-1) -- (mat-\aii-2);
	}
	\foreach \ai in {2,4,...,10,17}
	{\foreach \aii in {3,5,...,9,15}
		\draw[->] (mat-\ai-2) -- (mat-\aii-3);
	}

\foreach \ai in {3,5,...,9,15}
{\foreach \aii in {4,8,12}
	\draw[->] (mat-\ai-3) -- (mat-\aii-4);
}
	\end{tikzpicture}
	\end{adjustbox}
	\caption{Schematic illustration of NN model. The input layer consists of the distance vectors of the atom from its nearest neighbors. The output layer has 3 nodes with the values of the forces in each direction.} 
		\label{fig:schemNN}
\end{figure}
In this section, we describe the structure of the DL model, as well as its input and output. 

{\bf Network architecture.} For each atom, the output of the model is a 3D vector of the Cartesian forces. For each atom, we find the $N$ nearest neighbors (we used $N=12$ in most simulations); the convergence of errors with respect to $N$ is discussed later in the text. We sort the atoms according to their distance (closest is first) and then assign for each neighbor the following quantities: (Z$_n$, $\mathbf{d}_n$, 1/$d_n$), where Z$_n$ is the atomic number of the n$^{th}$ neighbor, $\mathbf{d}_n$ is the distance vector, and 1/$d_n$ is the reciprocal of the scalar distance. In this work we only analyze mono-atomic systems, so the number Z$_n$ is irrelevant and does not contribute to the model.

The input is fed to a fully connected neural network with two hidden layers. The first hidden layer has $L$ nodes and the second hidden layer has $L/3$ nodes. $L$ was typically around $3800$ for Si and around $250$ for Al and Na, we analyze the model MAE sensitivity as a function of $L$ later in the text. Finally, the output corresponds to the 3 Cartesian forces. We used the CRELU function for activation of the two first layers and a linear function for the output layer. We tried more hidden layers and larger hidden layers, but found that this did not improve much the model accuracy. A typical model architecture is illustrated in Figure \ref{fig:schemNN}

{\bf Network training and optimization} We used the Adaptive Moment Estimation (Adam) \cite{Kingma2014}
algorithm to train the model against the results of quantum calculations for the forces. We used a learning rate of 0.0001 and mini-batches of 100 samples.

\section{Results}

\subsection{Model performance for aluminum, sodium and silicon}

\begin{figure*}[t]
	\centering
	\includegraphics[width=1.0\linewidth]{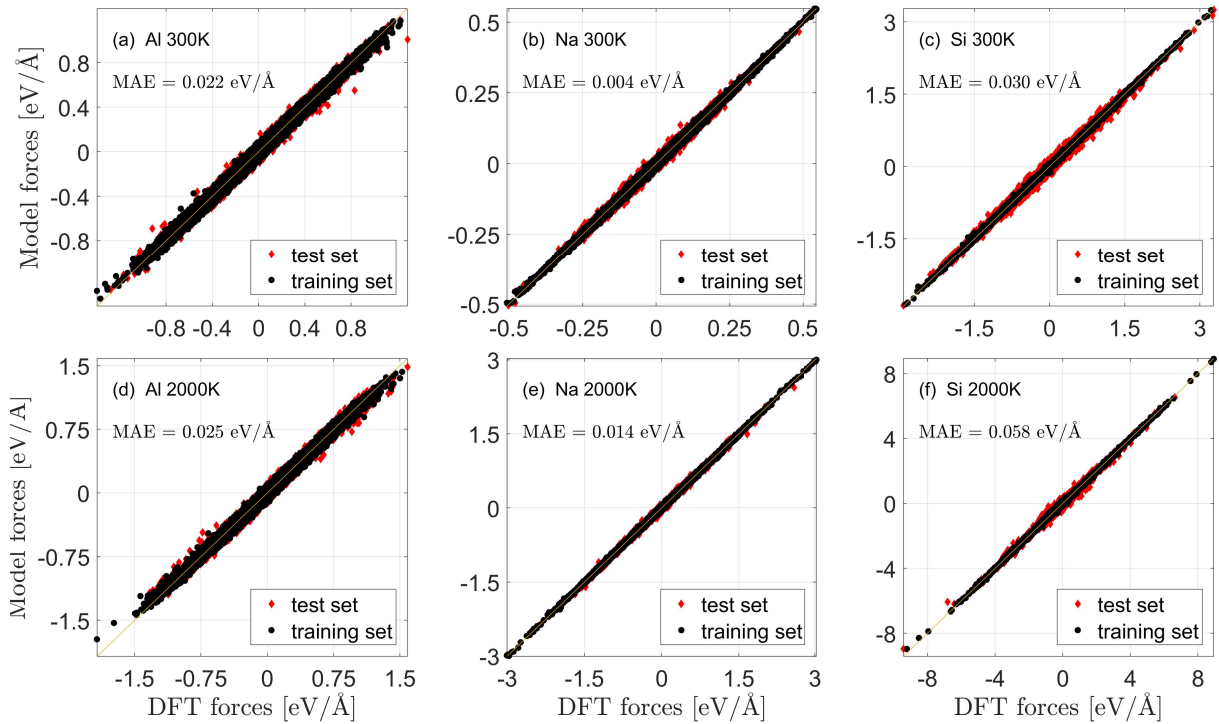}
	\caption{Comparison of the estimated forces to DFT forces at 300K ((a)-(c)) and 2000K ((d)-(f)). Results are shown for Na (27 atoms, (a),(d)); Al (27 atoms (b),(e)); and Si (16 atoms, (c),(f)). The training set results are shown with black dots while the test set results are shown with red dots. The solid line is a result of linear fit between the model and DFT results.}
	\label{fig:fig2}
\end{figure*}

In this section, we show the model performance when the cell and temperature are the same for the training and validation of the model. We used cells of 27 atoms for Al and Na and 16 atoms for Si. The cells were trained and tested at 300K and 2000K. The results are shown in Figure \ref{fig:fig2} and demonstrate that a small enough mean average error (MAE) was achieved.  The MAEs are also listed in Table \ref{tab:MAEComp}.
In all cases, 12 neighbors were used for input.

\begin{table}[h]
	\centering
	\begin{tabular*}{\linewidth}{@{\extracolsep{\fill}}p{0.1\linewidth}p{0.25\linewidth}p{0.09\linewidth}p{0.09\linewidth}p{0.12\linewidth}p{0.17\linewidth}@{}}
		\hline
		Atom&Training temperature [K]&Unit cell size&K points&MAE this work [ev/\AA]&MAE   literature [ev/\AA]\\
		\hline
		Al&2000&27 &125&0.025&0.02\cite{Botu2015c}\\
		&300 &27 &125&0.022 	&0.01\cite{Huan2017} 	 \\
		\hline
		Na&2000&27 &125&0.014& 	 \\
		&300 &27 &125& 0.004	& 	 \\
		\hline
		&2000&16 & 8 &0.058	&0.08\cite{Huan2017}\\
		Si&300 &16 & 8 &0.030 	&0.1\cite{Li2015,Glielmo2017} 	 \\
		\hline
	\end{tabular*}
	\caption{MAE comparison for all the systems. In all of the cases in this table the temperature and the cell size are the same in the training and the validation sets.}
	\label{tab:MAEComp}
\end{table}

A comparison of these results with those from previous studies shows that the presented DL models reach sufficient accuracy for the atomic forces. Such an accuracy was shown to allow running MD simulations without any pre-calculated FF at an accuracy that is close to that of AIMD~\cite{Han,Shakouri}.
It should be noted that the inclusion of the $1/d_n$ term was important to achieve reasonable MAEs. The inclusion of additional terms such as $1/d_n^6$ and $1/d_n^{12}$ did not help to further improve the results. 

\subsection{Sensitivity to the number of neighbors}
\begin{figure}[!htb]
	\centering
	\includegraphics[width=0.7\textwidth]{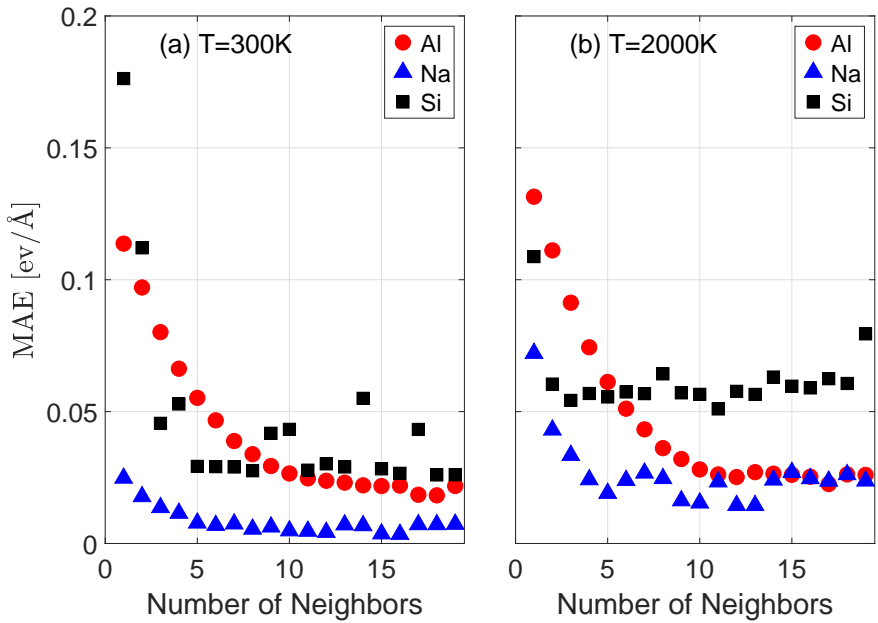}
	\caption{Mean Absolute Error (MAE) as function of number of input neighbor atoms that are used for the model (data produced at 300K (a) and at 2000K (b)). Al is shown with red circles, Na with blue triangles and Si with black squares.}
	\label{fig:fig3}
\end{figure}

\begin{figure*}
	\centering
	\includegraphics[width=1.0\linewidth]{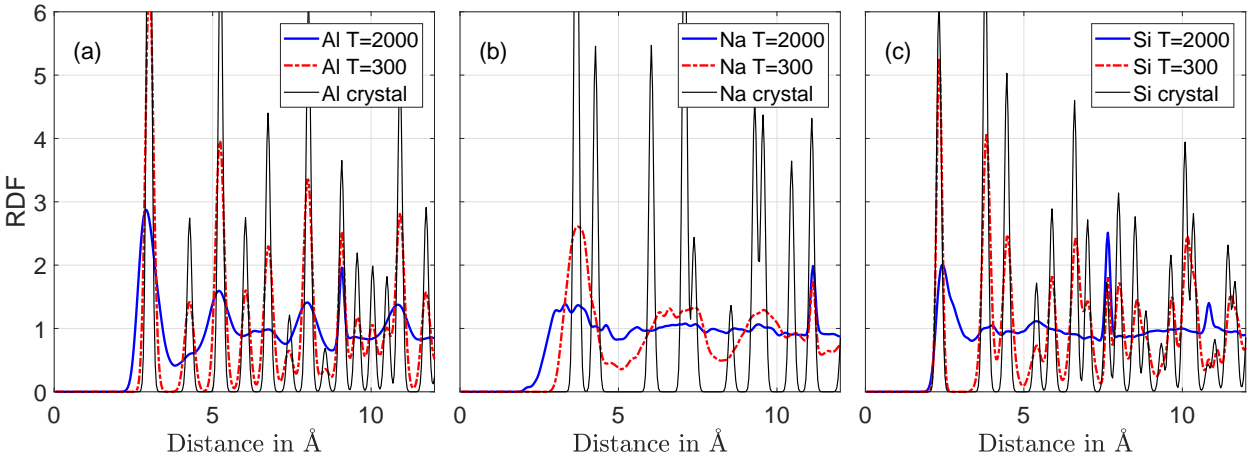}
	\caption{Average Radial Distribution Function (rdf) of the different structures - (a) Al, (b) Na and (c) Si. The rdf of the ground state crystal is shown with a solid black line, the rdf at T=300K is shown with a dash-dot red line and the T=2000 rdf is shown with a solid blue line. }
	\label{fig:figrdf}
\end{figure*}

In this section we analyze the MAE dependence on the number of neighbor atoms that were used in the model. We performed this analysis both at 300K and 2000K. At 300K all of the three materials are solids.  The melting point for Si/Al/Na is 1687K/933K/371K, respectively~\cite{lide2003crc}. Since we conducted the simulations at a constant volume this is an underestimation and we can assume that the materials are somewhere between solid and liquid. A possible measure of the material atomic structure is the radial distribution function (rdf). Figure \ref{fig:figrdf} shows the normalized rdf of the different systems at both 300K and 2000K. At 300K, both Si and Al exhibit an rdf that is close to the crystalline system. In contrast, while for Na it still shows peaks that are related to the crystalline structure, it is already heavily smeared. At 2000K, all systems are heavily smeared, with Al still showing some structural peaks. It is very clear that at 2000K there are many distances that do not appear at 300K.

We expected that systems with more nearest neighbors, like Al (fcc, 12 n.n.), will generally require more neighbors to converge in comparison to systems like Na (bcc, 8 n.n.) or Si (diamond, 4 n.n.). Furthermore, we expected that this trend will be clearer at the lower temperature where all the materials are solids. Figure \ref{fig:fig3}a shows the results for 300K and \ref{fig:fig3}b for 2000K. It is evident that at both temperatures Si and Na converge faster than Al. Si converges slightly faster than Na at 2000K, but reaches a significantly higher converged MAE. The trends are in fact clearer at the higher temperature; one possible reason is that the AIMD at that temperature covered a wider range of configurations. Further analysis of this trend will follow in future work.  

This analysis, namely of MAE as a function of n.n., makes it possible to uncover physical attributes of the system in question from the DL algorithm.
While in the systems we analyzed, the physical attributes are known, we suggest that for more complex systems, such an analysis can give new insight to the internal structure of the system.

\subsection{Temperature analysis}

\begin{figure*}
	\centering
	\includegraphics[width=1.0\linewidth]{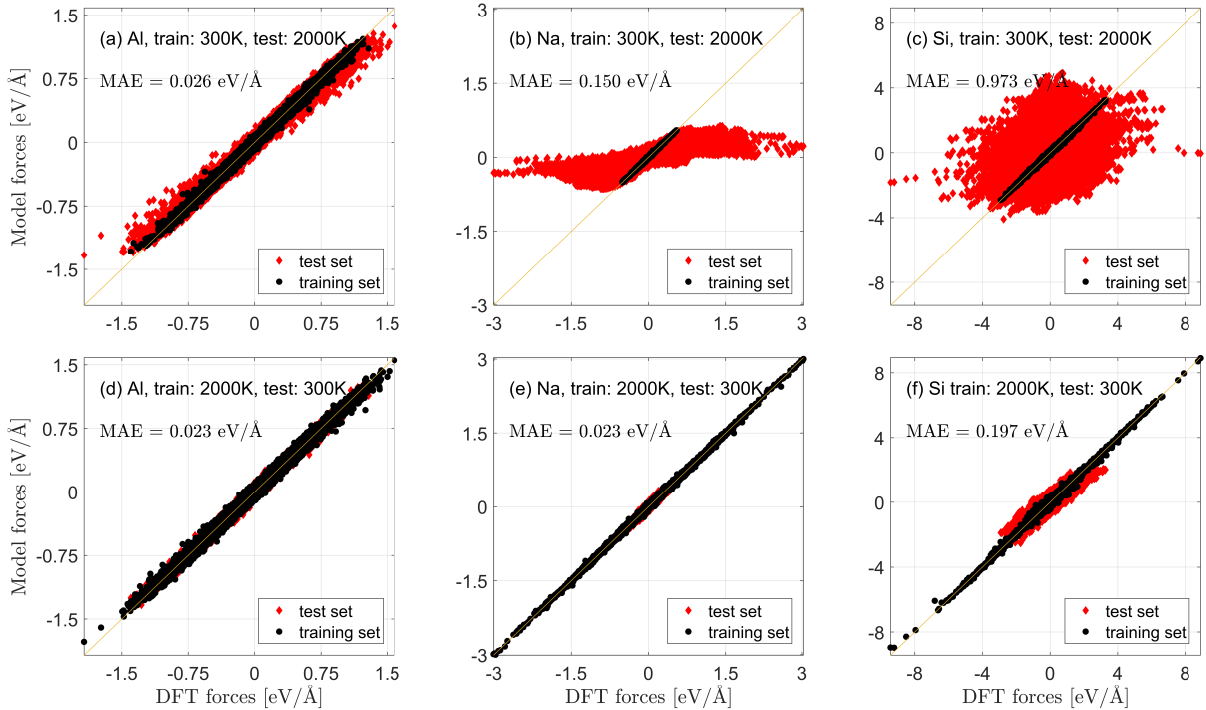}
	\caption{Comparison of the estimated forces to DFT forces at 300K ((a)-(c)) and 2000K ((d)-(f)). Results are shown for Na (27 atoms, (a),(d)); Al (27 atoms (b),(e)); and Si (16 atoms, (c),(f)). The training set results are shown with black dots while the test set results are shown with red dots. The solid line is a result of linear fit between the model and DFT results.}
	\label{fig:fig5new}
\end{figure*}

Here, we analyzed the ability of a model that was trained at one temperature to predict results at another temperature. Naturally, if we use low temperature MD as our training set, there is a high chance that most of the set is around the ground state minimum energy. Therefore, the outcome might not predict well for other meta-stable minima of the potential energy surface. Furthermore, as is obvious from Figure \ref{fig:figrdf}, the T=300K simulation does not always deviate enough from the ground state crystal, and so some distances that exist at T=2000K are completely absent at T=300K. If we use high temperature MD results as our training set we have a higher chance to cover more configurations, but we might have a more sparse coverage for each minimum. Consequently, it is natural to assume that training at higher temperatures might lead to performing well at lower temperatures while the opposite is less probable. In Figure \ref{fig:fig5new} we show the results of all systems for 300K$\rightarrow$2000K and for 2000K$\rightarrow$300K. It is very clear from the figure that the first case - training at low and testing at high temperature - yields poor performance; this is especially true for Si, where it is clear that the model seems to be almost random. The second case - training at high temperature and testing at low temperature - gives a higher MAE in comparison to same temperature tests, but behaves reasonably well and can be used. This result shows that it is possible to construct temperature-transferable DL models for MD simulations.

\subsection{Scaling analysis}
\begin{figure*}
	\centering
	\includegraphics[width=1.0\linewidth]{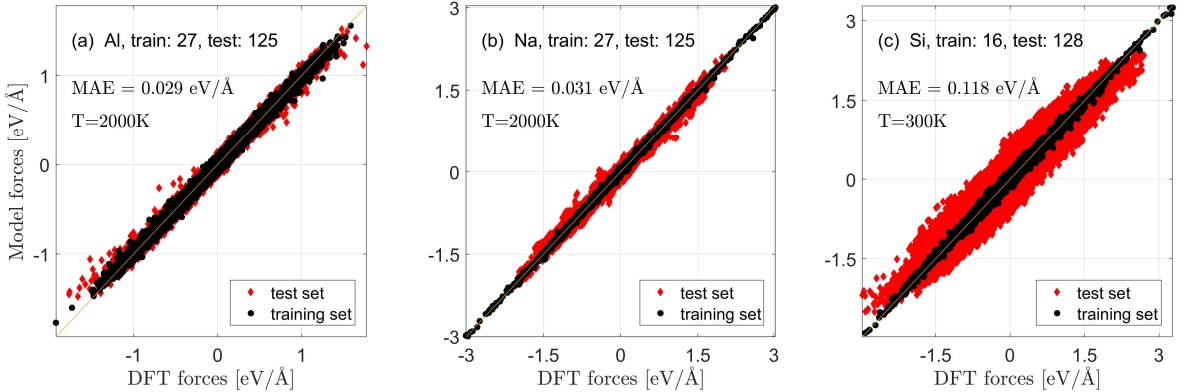}
	\caption{Comparison of estimated forces and DFT forces when the sizes of the training system and validation system are not the same. The following sets are shown: (a) Al, train with 27 atoms, validation with 125, (b) Na, train with 26, test with 125, (c) Si, train with 16, test with 128. The training set results are shown with black dots while the test set results are shown with red dots. The solid line is a result of linear fit between the model and DFT results.}
	\label{fig:fig6new}
\end{figure*}

In this section we evaluate the ability of a model that was trained with a cell of a given size to perform with cells that are larger. As we use a local environment for the training input, the model is, in a way, `blind' to the number of the atoms in the cell. Clearly, there can be long-range forces that a \dblquote{local} environment will not capture. A similar problem can also exist in classical FF that do not include polarization terms and have a cutoff. A full solution to the problem of scaling will require some specific treatment or training for the response to long-range forces and is beyond the scope of this work. We can still hope that the local model can yield reasonably good force predictions in many scenarios. In Figure \ref{fig:fig6new} we show the prediction of forces for the three materials. For Al (2000K), training with 27 atoms and testing with 125 yields an MAE of $\sim$0.03eV/\AA, which is comparable with the same size performance. For Na (2000K), training with 27 atoms and testing with 125 atoms yields an MAE of $\sim$0.03 eV/\AA, which is a bit worse than the same size model performance. For Si (300K), going from 16 atoms to 128 atoms yields an MAE of 0.12 eV/\AA, higher than the same size performance of 0.03eV/\AA.

It is evident that reasonable scaling was demonstrated for Al and Na, which means that we could use the smaller cell to estimate errors in the larger cell. With Si, further work should be done, as the performance penalty is a bit too high.
\section{Network architecture analysis}
\begin{figure}
	\centering
	\includegraphics[width=0.7\linewidth]{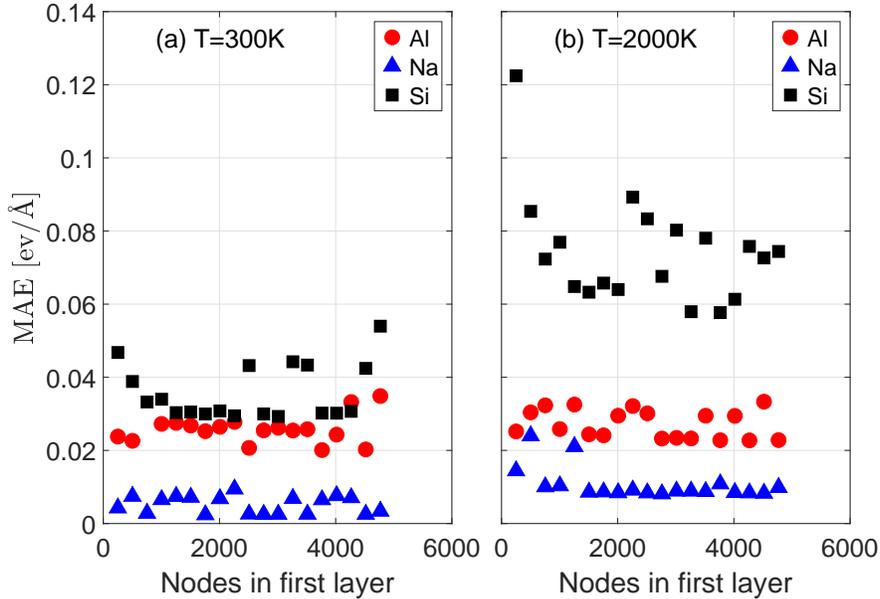}
	\caption{MAE dependence on the number of nodes, $L$, of the first hidden layer: (a) showing T=300K, (b) showing T=2000K. Al is shown with red circles, Na with blue triangles, and Si with black squares.}
	\label{fig:fig7new}
\end{figure}

To study the required network size, we checked the MAE's sensitivity to the number of hidden layers and the size of the first layer. In the first test, we used two hidden layers, the first with $L$ nodes and the second with $L/3$ nodes, and we varied $L$. This parameter can strongly affect the model computational efficiency. Since we have two layers, we can expect the model computation time to have $\mathcal{O}(L^2)$ scaling. Figure \ref{fig:fig7new}a demonstrates that for T=300K, In Na and Al, we can reduce $L$ to $\sim250 \simeq 20N$ ($N=12$ being the number of neighbors that are used) without significantly increasing the MAE. With Si, even at T=300K, there is an improvement in performance when increasing $L$ to $\sim2000$. The picture at T=2000K, shown in Figure \ref{fig:fig7new}b, is slightly different. For Si, we see significant but slow improvement in performance when increasing $L$, however, for $L$ above 2000 the algorithm starts to have convergence problems and does not always find the minimal possible MAE. A larger L means more degrees of freedom for the model and hence theoretically lower MAE, in practice, at some point there are too many degrees of freedom for a given set of data and hence convergence becomes more difficult. In Si and Na, it is evident that the improvement with $L$ is more significant at the higher temperature. An essential physical reason for the ability to use small $L$ at low temperatures is the following. As the temperature becomes low, the deviations from equilibrium can be described mostly within the harmonic approximation. Hence, the calculated quantum forces become linear with the distance vector. A description of a linear transformation of the distances requires a minimal $L$ and so it is easy to build a small model for the forces prediction. As the temperature becomes high, there is a significant deviation from the harmonic approximation, and hence the energy function becomes more complicated. This more complicated energy surface requires more domains of piecewise linearity for the forces and therefore a larger $L$. 

\begin{figure}
	\centering
	\includegraphics[width=0.7\linewidth]{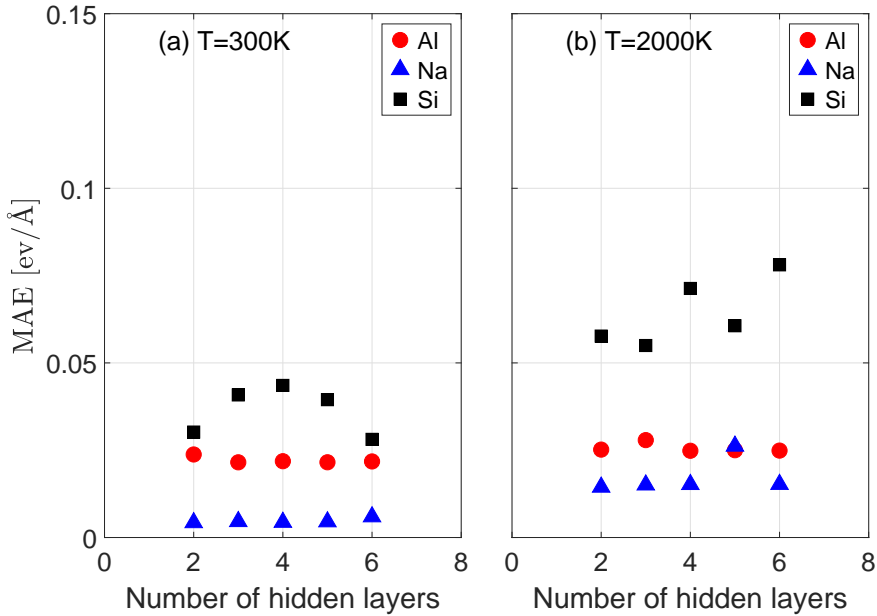}
	\caption{MAE dependence on the number of hidden layers. The x-axis is the number of hidden layers while the y-axis is the model's MAE: (a) T=300K, (b) 2000K. Al is shown with red circles, Na with blue triangles, and Si with black squares.}
	\label{fig:fig8new}
\end{figure}

In the second test, we used $L\simeq 200N$ and checked whether increasing the number of hidden layers helps to improve the results. In this test, the first hidden layer was with $L$ nodes and all the next layers had $L/3$ nodes. As is evident from Figure \ref{fig:fig8new}, increasing the number of hidden layers beyond two does not improve the error. This trend is true for both T=300K and T=2000K. Increasing the number of hidden layers beyond 6 resulted in over-fitting problems, probably because more data was needed for the amount of parameters that are fitted.

\section{Summary and Discussion} 
In this work, we presented and implemented a new DL model for atomic forces. This model enables predicting directly atomic forces with a close to DFT accuracy while using a MD sampling learning procedure. We also studied the model properties and the physical insights that this model provides.

DL and ML models differ from classical FF by not assuming an explicit physical model for the forces. We can therefore expect that their transferability behavior from one temperature to another might be different as well. In this work, we used MD as the sampling method to construct the training and test set. We showed that sampling at low-temperatures does not have sufficient coverage of the configuration space and so the trained model cannot predict the behavior at a higher temperature. 
In contrast to low-temperature sampling, once we use a training set constructed from high-temperature MD runs, we can use the model at lower temperatures with reasonable accuracy. This can be explained by a better sampling of the configuration space at higher temperatures. We expect that other DL and ML models will behave similarly, as the lack of any assumption of an explicit physical model for the forces is common to most of them. 

Another significant observation is the scalability property of the suggested model. As we showed above, for Al and Na, one can study relatively small systems of size 3x3x3 unit cell (27 atoms) to gain knowledge on much larger 5x5x5 unit cells (125 atoms). We demonstrate this in Figure \ref{fig:fig6new}, which shows results for the transition from small to large cells. For Al and Na, this was shown for data sets that were produced at 2000K, which can have forces that are significantly beyond the harmonic regime. 

The analysis for the model sensitivity to the number of neighbors that are used, as shown in Figure \ref{fig:fig3} for T=300K and T=2000K, demonstrates that we can learn some physical properties of the system (e.g., the atom coordination) from the model performance. Some qualitative trends are evident from the graphs; first, at low temperatures, even the first neighbor can produce a reasonable estimation for the forces in the metals Al and Na. However, Si, which has four covalent bonds, needs at least four neighbors at both checked temperatures to estimate reasonably the atomic forces. At the high temperature, more than 12 neighbors are required for Al, and more than 6 neighbors are needed for Na, where 12 and 8 are the numbers of first nearest neighbors for the fcc Al and bcc Na structures.

To summarize, the presented model demonstrates two essential properties: size scalability and temperature transferability. Size scalability means that one can predict atomic forces of large systems while learning from small systems that one can study by DFT. Temperature transferability of a force field is extremely important for MD simulations when one would like to find phase transitions and temperature dependent processes. Obviously, there are also limitations in this method; first of all, a \dblquote{local} predictor will probably underperform in situations where long-range forces, not captured by the model, dominate the picture. Furthermore, it is evident that the performance is not equally good for different materials, with Si seen to be more challenging, and this might require a more complicated model. Finally, we have checked a relatively homogeneous environment and, while the environment is \dblquote{local}, there might be a need for significant additional training in situations that include interfaces and surfaces. 

\begin{acknowledgement}
AN acknowledges support from the Pazy foundation (grant 281/18) and the Planing \& Budgeting Committee of the Council of High Education and the Prime-Minister Office of Israel, in the framework of the INREP project. GG acknowledges support by the Jacobs Foundation.
\end{acknowledgement}

\providecommand{\latin}[1]{#1}
\makeatletter
\providecommand{\doi}
{\begingroup\let\do\@makeother\dospecials
	\catcode`\{=1 \catcode`\}=2 \doi@aux}
\providecommand{\doi@aux}[1]{\endgroup\texttt{#1}}
\makeatother
\providecommand*\mcitethebibliography{\thebibliography}
\csname @ifundefined\endcsname{endmcitethebibliography}
{\let\endmcitethebibliography\endthebibliography}{}

\end{document}